\documentclass[10pt,a4paper]{article}
\usepackage[utf8]{inputenc}
\usepackage{amsmath}
\usepackage{amsfonts}
\usepackage{amssymb}
\usepackage{graphicx}
\usepackage{float}
\usepackage{url}
\usepackage[left=2cm,right=2cm,top=2cm,bottom=2cm]{geometry}
\begin{document}
\begin{figure}[b]
\includegraphics[width=1\textwidth]{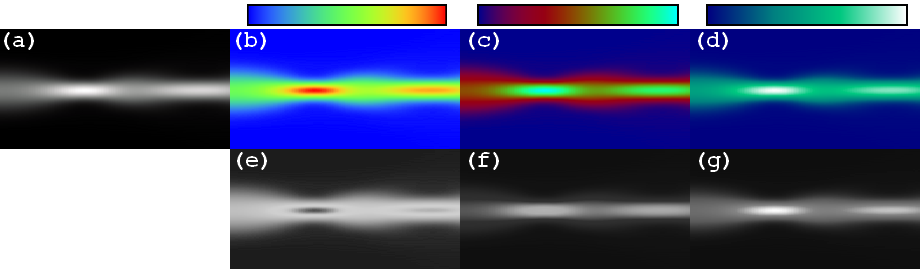}
\caption{(a) Soliton propagation (original image), (b) image presented with a common (rainbow) false color palette, (c)-(d) images encoded with proposed palettes (P\#2 and P\#3), (e)-(g) luminance of the false color images.}
\label{fig:soliton}
\end{figure}
\begin{figure}[]
\includegraphics[width=1\textwidth]{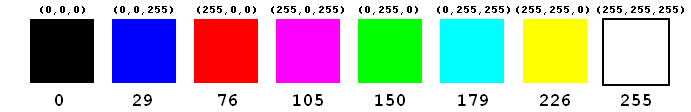}
\caption{Basic colors and their luminance (bottom) and RGB channels (top).}
\label{fig:palette_colors}
\end{figure}
\begin{figure}[H]
\includegraphics[width=1\textwidth]{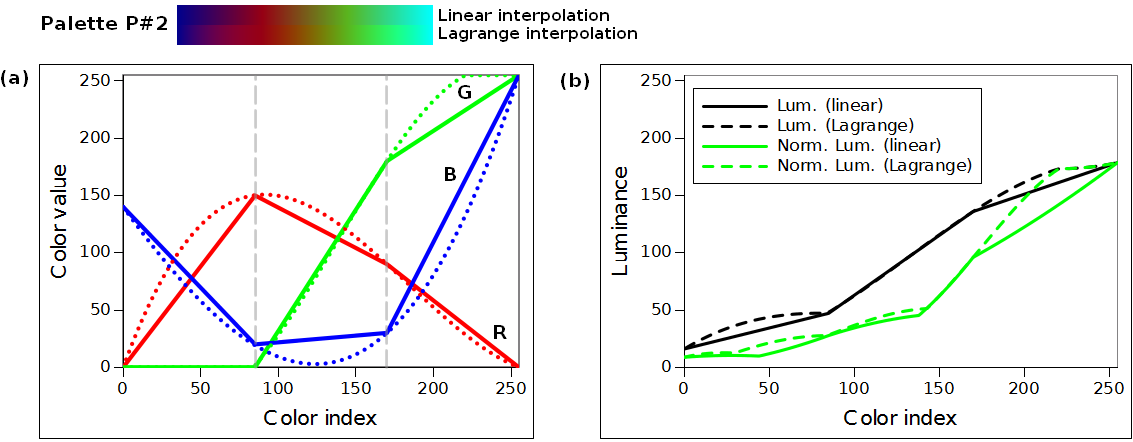}
\caption{(a) Color value of the red, green and blue channels as a function of color index. Four basic points are interpolated with line functions (solid line) and with Lagrange polynomials (dotted line). (b) Luminance and normalized luminance of the palette.}
\label{fig:luminance_value}
\end{figure}

 \begin{center}
  \Large 
  \textbf{Design of false color palettes for grayscale reproduction}\\
  
  \vspace{15pt} 
  \large  
  Filip A. Sala\\
  \vspace{25pt}
  \normalsize   
  Warsaw University of Technology, Faculty of Physics, Koszykowa 75, 00-662 Warsaw, Poland
  \vspace{25pt} 
 \end{center}
\small 
\textbf{Abstract} Design of false color palette is quite easy but some effort has to be done to achieve good dynamic range, contrast and overall appearance of the palette. Such palettes, for instance, are commonly used in scientific papers for presenting the data. However, to lower the cost of the paper most scientists decide to let the data to be printed in grayscale. The same applies to e-book readers based on e-ink where most of them are still grayscale. For majority of false color palettes reproducing them in grayscale results in ambiguous mapping of the colors and may be misleading for the reader. In this article design of false color palettes suitable for grayscale reproduction is described. Due to the monotonic change of luminance of these palettes grayscale representation is very similar to the data directly presented with a grayscale palette. Some suggestions and examples how to design such palettes are provided. \vspace{10pt} \\ 
\textbf{Keywords:} false-color palettes; data visualization;\vspace{10pt}\\
\normalsize
False color palettes are used to display information in variety of fields like physical sciences (e.g. astronomy, optics)\cite{NASA_color,Netravali_book, aguilar2014paintings}, medicine (e.g. in different imaging systems)\cite{vardasca2014medicalimages} and industry \cite{toet2012nightvision}. In most applications palettes are optimized for better contrast, dynamic range and overall appearance of the palette \cite{Orchard_colorquantization, stauffer2015meteorological}. They are adjusted for particular data that have to be presented. There is no unique algorithm or easy rules to make a good false color palette. However, most of them make the data easier to interpret and details easier to see, the most commonly used "rainbow" palette is a subject of criticism. Some works point that it obscures the data and 
may produce artifacts \cite{Borland_rainbowcolormap,EndofRainbow2004}. Moreover most of commonly used palettes printed in grayscale or displayed on a monochromatic e-ink reader leads to ambiguous mapping and misleading artifacts. In this article some rules and algorithms how to prepare appropriate color palette which will also look good after conversion into grayscale are presented. At the end some exemplary palettes ready to use are shown. 
In Fig. \ref{fig:soliton}a propagation of an optical beam (soliton) \cite{karpierz6,sala_OptExpress2012, sala_jnopm2014} in a nonlinear medium is presented. It is clearly visible that after converting the data encoded with rainbow palette into grayscale (Fig. \ref{fig:soliton}e)  the image becomes confusing. Dark colors represent both the lowest and the highest beam intensities. For the proposed palettes (Fig.\ref{fig:soliton}f,g) the grayscale representation looks much more similar to the original image (Fig. \ref{fig:soliton}a). 
The main idea is to achieve the luminance to be monotonically increasing or decreasing with the color index of the palette. Each color in the palette has its index $i$. For instance to get monotonically increasing luminance of the palette the luminance of the $i+1$ color has to be greater than luminance of the $i$ color. According to ITU BT.601 recommendation \cite{ITU601} luminance can be defined as 
\begin{equation}
L_i = 0.299 \,R_i + 0.587 \,G_i + 0.114\, B_i
\end{equation}
where $i$ - is a color index and $R_i,G_i,B_i$ represent value of the red, green and blue channels respectively. This is the most common and widely used definition of luminance in digital image processing. Thus it is likely that most software will use it to convert a color image into grayscale. The same also applies to printers, e-book readers and other popular devices. Now let's consider 24-bit color space (8 bits per channel), so the maximum value of the color is 255. In this color space there are a few basic colors presented in Fig. \ref{fig:palette_colors} with their corresponding luminance. Now it is even more clear that rainbow palette with red representing high values, blue representing low values and green representing middle values can not have monotonically changing luminance. 
\noindent
Starting from these 8 basic colors it is easier to build an appropriate palette. To design a palette a few main colors have to be chosen and then interpolated. At first let us consider a palette based on 4 points having two end points at $i=0$, $i=255$ and two middle points at $i=85$, $i=170$. The luminance of these colors have to fulfill the following relation $L_0 < L_{85} < L_{170} < L_{255}$. Subsequently these colors have to be interpolated for instance with a line function for each channel (red, green, blue) separately. To obtain smoother color change Lagrange interpolation can be employed. Example of a palette interpolated using both methods is presented in Fig. \ref{fig:luminance_value}.
\noindent
When it comes to Lagrange interpolation of 4-point color palette it can be written as:
\begin{eqnarray}
\nonumber P(i)=-\frac{(i-85)(i-170)(i-255)}{3684750}\cdot c_0+\frac{i(i-170)(i-255)}{1228250}\cdot c_{85}\\-\frac{i(i-85)(i-255)}{1228250}\cdot c_{170}+\frac{i(i-85)(i-170)}{3684750}\cdot c_{255}
\end{eqnarray}
where $P(i)$ - color value at index $i$. Constants $c_0,c_{85},c_{170},c_{255}$ - are fixed color values at points $i=0, i=85, i=170, i=255$. This interpolation has to be done for all color channels i.e. red, green and blue.
For 3-point palettes Lagrange interpolation can be noted in a general form as:
\begin{equation}
P(i)=\frac{(i-m)(i-255)}{255\cdot m}\cdot c_0+\frac{i(i-255)}{m(m-255)}\cdot c_m+\frac{i(i-m)}{255(255-m)}\cdot c_{255}
\end{equation}
where $m$ is an index of the middle point. 
When the middle point $c_m$ is at index $m=127$ than it simplifies to:
\begin{equation}
P(i)=\frac{(i-127)(i-255)}{32385}\cdot c_0- \frac{i(i-255)}{16256}\cdot c_{127}+\frac{i(i-127)}{32640}\cdot c_{255}
\end{equation}
Lagrange polynomials can give values higher than maximum color value (in this example higher than 255) or negative values. It is important to fix this problem by substituting negative values with zero and values higher than 255 with 255. Both methods of interpolation will give real color values which have to be converted to integer values. No matter which function will be used, round, ceiling, floor etc. there might be some non-monotonic behavior of the luminance caused by this conversion. Hopefully this fluctuations are minor (typically luminance change is much less than 1) and are clearly not visible to a human eye.
 
\begin{figure}[]
\begin{center}
\includegraphics[width=0.5\textwidth]{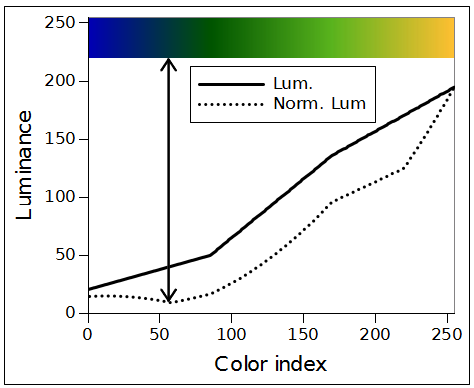}
\caption{Luminance of the color palette with monotonically increasing luminance (solid line) but with non-monotonic normalized luminance (dotted line). Darker area in the palette is denoted with an arrow.}
\label{fig:palette5}
\end{center}
\end{figure}

\begin{figure}[]
\begin{center}
\includegraphics[width=0.89\textwidth]{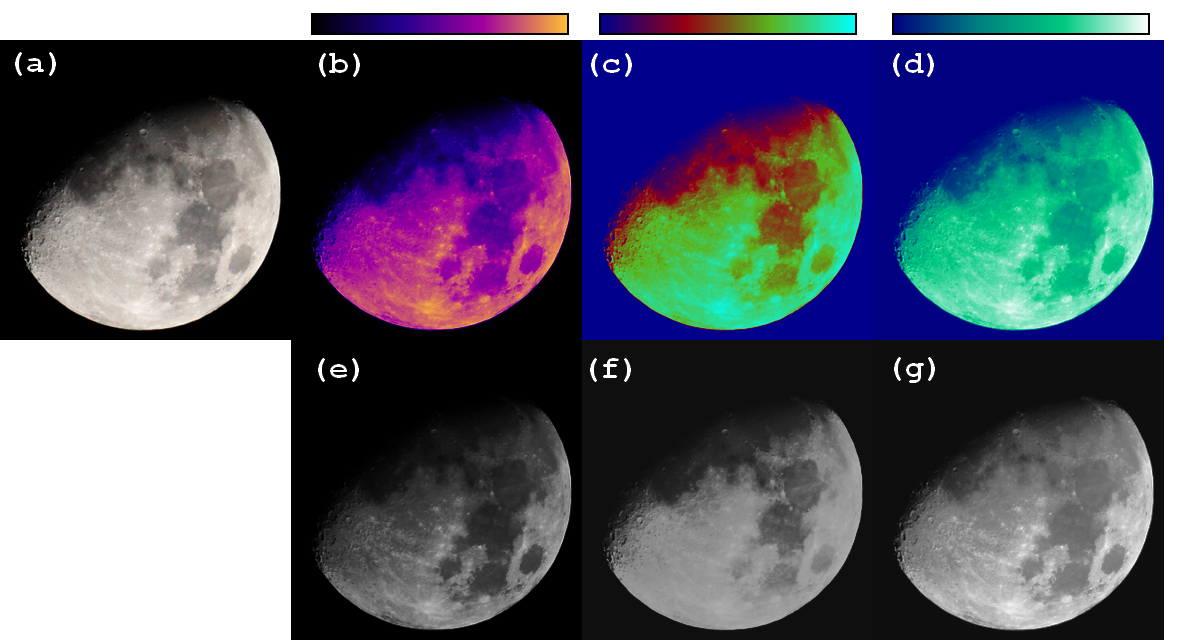}
\caption{(a) Original image of the Moon, (b)-(d) false color images (palettes P\#1, P\#2 and P\#3 were used), (e)-(g) luminance of the false color images. Palettes were interpolated with a line function.}
\label{fig:moon}
\end{center}
\end{figure}

However, luminance is a quite good parameter to describe the brightness as it reflects response of the human eye to different colors it is not perfect. It is possible to design a color palette which luminance is quickly increasing but the palette appears to have darker parts which leads to unnatural look and artifacts in the presented data. Such palette is presented in Fig. \ref{fig:palette5}. Thus another parameter, which has to be taken into account, is a normalized luminance which can be defined as:
\begin{equation}
L_i^\prime=L_i \cdot V_i / 255 \quad \text{where} \quad V=\text{Max}(R_i,G_i,B_i)
\end{equation}
This parameter also should monotonically increase with color index. Each local minimum leads to darker areas in the palette. 
False color palettes are commonly used in astronomy and image processing fields. It appears that palettes interpolated with a line function are better for images as they typically have quicker changes in luminance (than Lagrange polynomials) which leads to better visibility of details on the image. On the other hand while presenting simulation results smoother changes are suitable and better reflect the presented data. Examples of the proposed color palettes interpolated with a line function are presented in Fig. \ref{fig:moon}.

\begin{figure}[t]
\begin{center}
\includegraphics[width=0.5\textwidth]{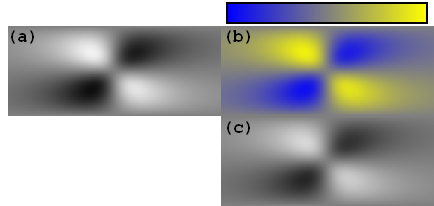}
\caption{(a) Molecular reorientation (original image), (b) false color image, (c) luminance of the false 
color image.}
\label{fig:molecular}
\end{center}
\end{figure}

In many cases there is a need to present data that reflect positive and negative values. To distinguish the sign of the presented data a color palette based on 3 points can be used. The middle point has to reflect zero value and can be a gray color for instance RGB$=(127,127,127)$. The end point colors must have very low luminance for low data values and very high luminance for high data values. It is also a good idea to choose corresponding colors as end points, e.g. blue and yellow, red and cyan etc. which give a line (affine) characteristic of the palette. However, it has to be avoided to choose colors which luminance is similar like magenta and green. An example of 3-point palette is presented in Fig. \ref{fig:molecular} which shows molecular reorientation of the liquid crystal \cite{sala2012_JOSAB}. For such purposes also a simple grayscale palette is very efficient. 

\begin{figure}[]
\begin{center}
\includegraphics[width=0.9\textwidth]{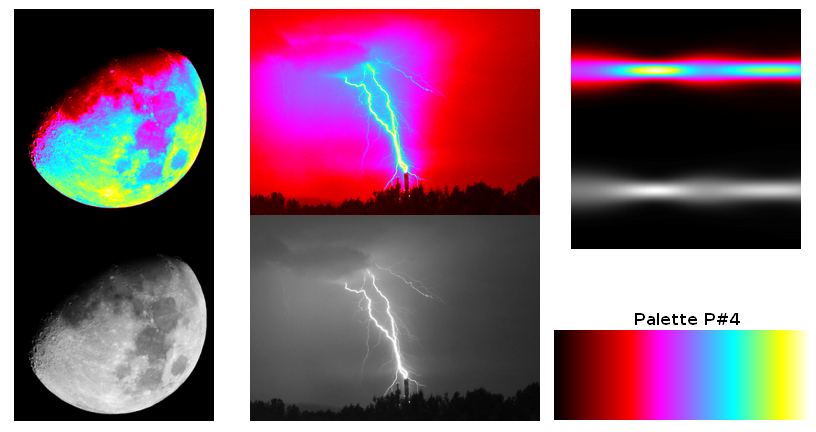}
\caption{False color images encoded with a linear palette based on 6 interpolation points (top) and corresponding luminance (bottom). As the palette is linear the luminance and the original grayscale image are the same.}
\label{fig:6point_comparison}
\end{center}
\end{figure}

\begin{figure}[]
\includegraphics[width=\textwidth]{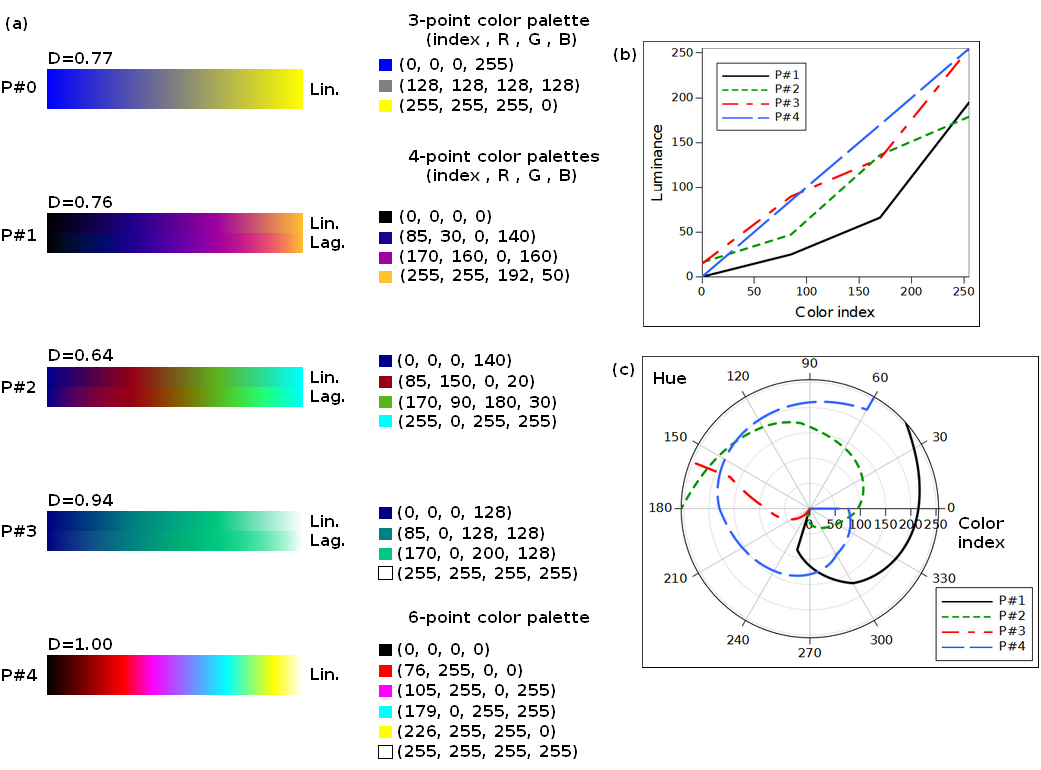}
\caption{Comparison of all palettes presented in the article. (a) color palettes interpolated with line function (Lin.) and Lagrange polynomials (Lag.) and information about the index and RGB channels of the interpolation points. $D$ denotes dynamic range of the palette. (b) Luminance of the palettes, (c) Hue of the palettes.}
\label{fig:palletes_comparison}
\end{figure}

However, described palettes were based on 3 or 4 color points it is obviously possible to make a palette with many points. In such case some difficulties may arise. First of all the interpolation with a Lagrange polynomial might not be effective anymore because of the high order of the polynomial. The higher the order the higher the possibility of local minima and unexpected behavior of interpolating function. Still line function or cubic spline interpolation or similar can be used. Another issue is a very quick change of hue because of many colors present in the palette. In some cases it might lead to artifacts. For instance an area on the image where luminance changes only slightly will be encoded with completely different colors. These are some minor issues which have to be managed in the process of designing a palette. A palette based on more colors can be easily designed by using basic colors from Fig. \ref{fig:palette_colors}. In the following example 6 colors were chosen: black, red, magenta, cyan, yellow and white. Green and blue colors were omitted to prevent ambiguous mapping and dark areas in the palette. The 6 colors were used as interpolating points at indexes corresponding to their luminance (i.e. satisfying the relation $i = L_i$). Then points were interpolated using line function. Such palette is not only monotonic but also linear at any point. Examples of application for this palette are presented in Fig. \ref{fig:6point_comparison}. As the palette is linear it means that the luminance of the false color image and the original grayscale image are the same. As described above taking different colors as interpolating points and placing them at indexes corresponding to their luminance is another possibility of building color palette. This technique works better for high number of interpolating points which have to be properly chosen. Some effort has to be done to avoid dark areas in the palette and repeated colors.   
Another important parameter describing palette is a dynamic range which can be defined as:
\begin{equation}
D = (L_{max} - L_{min}) /255
\end{equation}
The higher the dynamic range the greater the overall difference in brightness of the palette. However it is important parameter for color palettes, it can be compensated with high change in hue. For instance, even if the palette has relatively low dynamic range (i.e. difference in luminance is low) but has wide range of hue the data still can be distinguished because of different colors. On the other hand for grayscale reproduction only the dynamic range counts. 
In figure \ref{fig:palletes_comparison} comparison of all color palettes used in this article is presented. Detailed information of interpolating points is also provided to allow the reader to easily reproduce the results. Dynamic range is also calculated for each palette and the graphs of luminance and hue are provided. Palettes with higher dynamic range will look better when reproduced in grayscale and those with higher change in hue (like P\#2 or P\#4 with hue range of about $300^\circ$) will better expose details of the presented image or data. All the presented results and suggestions also applies to the palettes with monotonically decreasing luminance. However, in such case the image converted into grayscale will be similar to the negative of the original data. It might be a good solution as background is encoded with a bright color which is better for printing. 
\paragraph{Conclusions} 
In this article discussion on false color palettes for grayscale reproduction was presented. It was proved that it is important to achieve monotonically increasing or decreasing luminance and normalized luminance versus color index. It was shown that palettes based on 3 interpolating points can be used to present the data with positive and negative values. Lagrange interpolation formulas were shown for palettes based on 3 and 4 points. It was also noticed that dynamic range and hue range are important parameters in the process of design. At the end a linear palette based on 6 interpolating points was shown. The presented palettes and suggestions are only a starting point for designing a good color palette. Different data and images, different fields of science and industry need different color palettes. Presented results and comments are to help the process of designing an appropriate palette for the particular application.

\end{document}